\documentclass[prl,twocolumn,aps]{revtex4}
\usepackage{epsfig}

\begin{document}
\begin{titlepage}

\title{Scaling Confirmation of the Thermodynamic Dislocation Theory} 
\author{J.S. Langer}
\affiliation{Kavli Institute for Theoretical Physics, Kohn Hall, University of California, Santa Barbara, CA 93106}

\author{K.C. Le}
\affiliation{Materials Mechanics Research Group and Faculty of Civil Engineering, Ton Duc Thang University, Ho Chi Minh City, Vietnam}\

\date{\today}

\begin{abstract}

We show that the thermodynamic dislocation theory (TDT) predicts a scaling relation between stresses, strain rates, and temperatures for steady-state deformations of crystalline solids, and that this relation is accurately obeyed by a wide range of experimental data for both aluminum and copper.  Unlike conventional phenomenological dislocation theories, the TDT is based on the second law of thermodynamics.  Its success implies that descriptions of solid deformation that are not based on the statistical mechanics of nonequilibrium processes cannot be relied upon to be predictive.  Thus there is an urgent need -- and a new opportunity -- to revitalize this central part of materials physics. 

\end{abstract}
\maketitle

\end{titlepage}

For almost a century, the dislocation theory of crystalline deformation has played a central role in materials science.  Unfortunately and very remarkably, this theory has made almost no progress for about fifty years.  Although crystalline solids are essential in engineering applications, and although modern physics-based experimental techniques have provided a wealth of information about the dislocations in these solids, the theories developed to explain dislocation-driven phenomena have become entirely phenomenological. They describe phenomena mathematically but do not explain them;  they are not predictive. In fact, some of them are manifestly incorrect.  

The cause of this theoretical failure seems clear.  Dislocation-driven deformations of solids are complex nonequilibrium processes involving macroscopic numbers of dynamical degrees of freedom.  Theoretical physicists know that they must use statistical methods to deal with such situations.  Especially important is the second law of thermodynamics, which states that driven complex systems must move toward their most probable configurations, i.e. that their entropies must be non-decreasing functions of time.  But leading materials scientists since the 1950's have asserted that dislocation energies are too large, and that dislocation entropies are too small, for the second law to be applicable. \cite{COTTRELL-53,COTTRELL-02} 

We have argued for a decade that those assertions are wrong.  The thermodynamic dislocation theory (TDT) is based directly on the second law.  It was introduced in 2010 \cite{LBL-10} and has been shown, in a series of publications since then \cite{JSL-15, JSL-16, JSL-17, JSL-17rev, JSL-18, JSL-PCH-19, LTL-17, LTL-18, Le-18, Le-19} to be capable of solving a wide range of the most important problems in solid mechanics including strain hardening, elastic-plastic yielding, shear banding, grain-size effects, and the like.  Those problems were out of reach of the conventional approaches.  But the question remains: How sure are we that the TDT is more reliable than the observation-based phenomenologies?  Can we use it confidently to solve important materials problems that have been left untouched by the conventional methods? 

To test the reliability of the TDT, we have used it to derive a scaling law for steady-state deformations.  Such scaling laws have been proposed in the past. For example, Kocks and Mecking devoted much of their definitive review article \cite{KOCKS-MECKING-03} to the search for scaling relations based on experimental data and phenomenological strain-hardening formulas. So far as we know, however, the TDT-based scaling relation is the first of these to be truly successful.  As we shall show, it is accurately obeyed over a wide enough range of experimental data to make it seem highly unlikely that there is anything fundamentally incorrect about it. Our increased confidence in the TDT now leads us to ask some questions that urgently need to be answered for both basic and applied reasons.   

The thermodynamic basis of the TDT has been presented in  earlier publications.  (See especially \cite{LBL-10,JSL-17rev,JSL-PCH-19}.)  Its main premise is that the dislocations in a deforming crystalline solid can be described -- indeed, {\it must} be described -- by an effective temperature $T_{e\!f\!f}$ that differs greatly from the ordinary, ambient temperature $T$.  $T_{e\!f\!f}$ is truly a ``temperature'' in the conventional sense of that word; it is  derived by invoking the second law of thermodynamics. It is also a true temperature in the sense that, as energy flows through an externally  driven system containing dislocations, effective heat is converted to ordinary heat and dissipated.  Thus, this driven, nonequilibrium system should be visualized as consisting of two weakly coupled subsystems: the dislocations at temperature $T_{e\!f\!f}$, and the rest of the system playing the role of a thermal reservoir at temperature $T$. 

For present purposes, we need to know only that, in steady-state shear flow, the areal density of dislocations is given by the usual Boltzmann formula:
\begin{equation}
\label{rhoss}
\rho_{ss} = {1\over a^2}\,\exp\,\Bigl[- {e_D\over k_B\,T_{e\!f\!f}^{ss}}\Bigr],
\end{equation}
where $a$ is a minimum spacing between dislocations, $e_D$ is a characteristic dislocation energy, and $T_{e\!f\!f}^{ss}$ is the steady-state effective temperature.  The quantity $k_B\,T_{e\!f\!f}^{ss}/e_D$, usually denoted by the symbol $\tilde\chi_{ss}$, is a measure of the degree of disorder of the subsytem of dislocations.  It is determined by the rate at which this subsystem is being driven, i.e. by the strain rate $\dot\epsilon$.  If that driving rate is slow enough that irreversible atomic rearrangements have time to relax before the strain has changed appreciably, then $\tilde\chi_{ss}$ must be independent of the strain rate.  Typical time scales for atomic motions are of the order of $10^{-10} s$.  Thus $\tilde\chi_{ss}$ must be a constant for strain rates up to $10^6/s$ or even higher.  It then follows from Eq.(\ref{rhoss}) that steady-state dislocation densities must also be constant across this range of driving rates, which includes most ordinary applications.  In \cite{LBL-10}, we used a Lindemann-like argument to estimate that $\tilde\chi_{ss} \sim 0.25$ which turns out to  be roughly correct. 

The second core ingredient of the TDT is the depinning (``double-exponential'') formula, which also is based on a comparison of time  scales.  We know that the dislocations in a deforming solid, under almost all circumstances, are locked together in an entangled mesh that can deform only via thermally activated depinning of pairwise junctions.  The depinning times are very much longer than the times taken by dislocation segments to jump from one pinning site to another.  Thus the depinning rate controls the deformation rate, and no other rates are relevant in this approximation. 
\begin{figure}[h]
\begin{center}
\includegraphics[width=\linewidth] {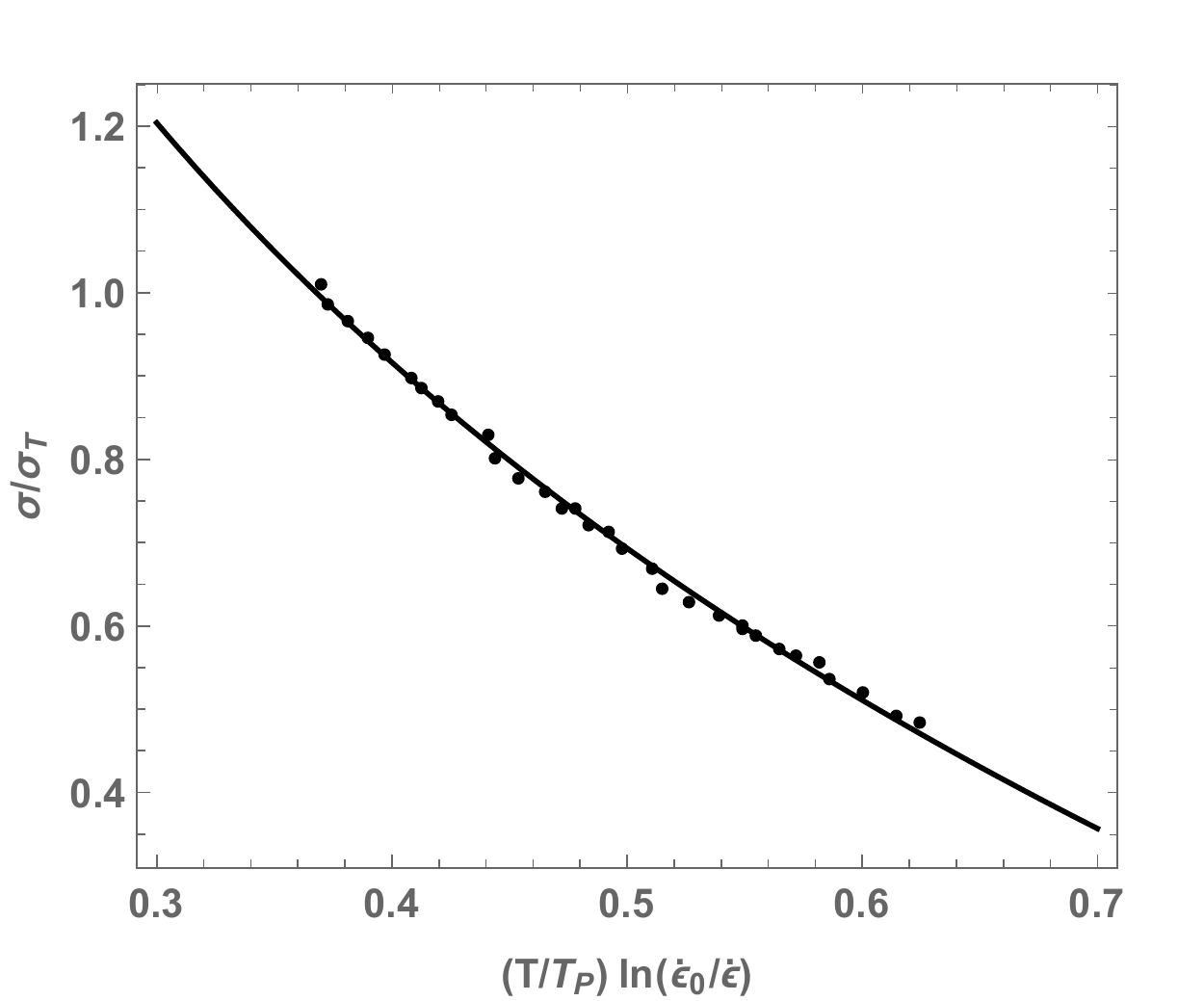}
\caption{Scaling relation given by Eq.(\ref{sigma}). The solid curve is the function $f(x) = \ln(1/x)$, with $x = (T/T_P)\,\ln(\dot\epsilon_0^{ss}/\dot\epsilon^{pl})$.  The data points are from \cite{SAMANTA-71} as interpreted in \cite{KCL-20}}   \label{TDTFig-1}
 \end{center}
\end{figure}
To be more specific, define the depinning rate to be $\tau_P^{-1} = \tau_0^{-1} \,\exp[- U_P(\sigma)/k_B\,T)] $, where $\tau_0$ is a microscopic time scale and $U_P$ is a pinning energy that depends on the applied stress $\sigma$.  Write this energy in the form $U_P(\sigma) = k_B T_P\,\exp[- \sigma/\sigma_T(\rho,T)]$, where $\sigma_T(\rho,T)$ is a characteristic stress that determines the magnitude of $\sigma$ necessary to reduce the pinning barrier by a factor of $1/e$.  If $a'$ is the separation between dislocations needed to produce this reduction, and $1/\sqrt{\rho}$ is the average distance between dislocations, then $a'\,\sqrt{\rho}$ is a strain, and $\sigma_T(\rho,T)=\mu(T)\,a'\,\sqrt{\rho}$ is a stress, where $\mu(T)$ is the (temperature-dependent) shear modulus.  In fact, $\sigma_T(\rho,T)$ is the Taylor stress.  To compute the plastic strain rate, use the Orowan formula $\dot\epsilon^{pl} = \rho\,b\,v$, where $b$ is the magnitude of the Burgers vector and $v$ is the average dislocation speed $1/(\tau_P\, \sqrt{\rho})$. The result is:
\begin{equation}
\label{doteps}
\dot\epsilon^{pl} = {b\over \tau_0}\sqrt{\rho}\, \exp\Bigl[- {T_P\over T} e^{- \sigma/\sigma_T(\rho,T)}\Bigr]
\end{equation}
or, equivalently,
\begin{equation}
\label{sigma}
{\sigma\over \sigma_T(\rho,T)} =  - \ln\,\Bigl[ {T\over T_P}\ln\Bigl({\dot\epsilon_0(\rho)\over \dot\epsilon^{pl}}\Bigr)\Bigr] 
\end{equation}
where  $\dot\epsilon_0(\rho) \equiv b\,\sqrt{\rho}/\tau_0$. 
 
For steady-state situations in which $\rho = \rho_{ss}$ remains constant, Eq.(\ref{sigma}) contains three system dependent but strain-rate independent parameters: $\sigma_T^{ss}\equiv \sigma_T(\rho_{ss},T)$, $\dot\epsilon_0^{ss}\equiv \dot\epsilon_0(\rho_{ss})$, and $T_P$. Thus, plots of measured values of $\sigma/\sigma_T^{ss}$ as functions of $(T/T_P) \,\ln (\dot\epsilon_0^{ss}/\dot\epsilon^{pl})$ should collapse onto a single curve once we have identified the values of those three parameters, which we can do by using known values of the modulus $\mu(T)$ and using a least-squares method to find the best fit between the parameters and  the scaling curve. 

To check this scaling hypothesis, we have used a set of compression  measurements by S.K. Samanta \cite{SAMANTA-71}.  These are old results, but they have the special advantage for us of using two different materials and testing them at different temperatures and strain rates under otherwise identical conditions.  Our scaling graph shown in Fig. 1 contains $32$ points: $12$ for pure copper at three temperatures and four strain rates, and $20$ for pure aluminum at four temperatures  and five strain rates.  Clearly, these points fall very accurately on the smooth curve predicted by the TDT analysis, which adds greatly to our confidence in this theory.

The time-dependent TDT consists of three physics-based equations of motion.  The first is Hook's law with the (``hypo-elasto-plastic'') assumption that elastic and plastic shear rates are additive:
\begin{equation} 
\dot \sigma = 2\,\mu (1+\nu)\,(\dot\epsilon^{tot} - \dot\epsilon^{pl}),
\end{equation}
where $\nu$ is Poisson's ratio and $\dot\epsilon^{tot}$ is the total elastic plus plastic strain rate.  $\dot\epsilon^{pl}$ is given by Eq.(\ref{doteps}), making this a highly nonlinear equation.  

\begin{figure}[h]
\begin{center}

\includegraphics[width=\linewidth] {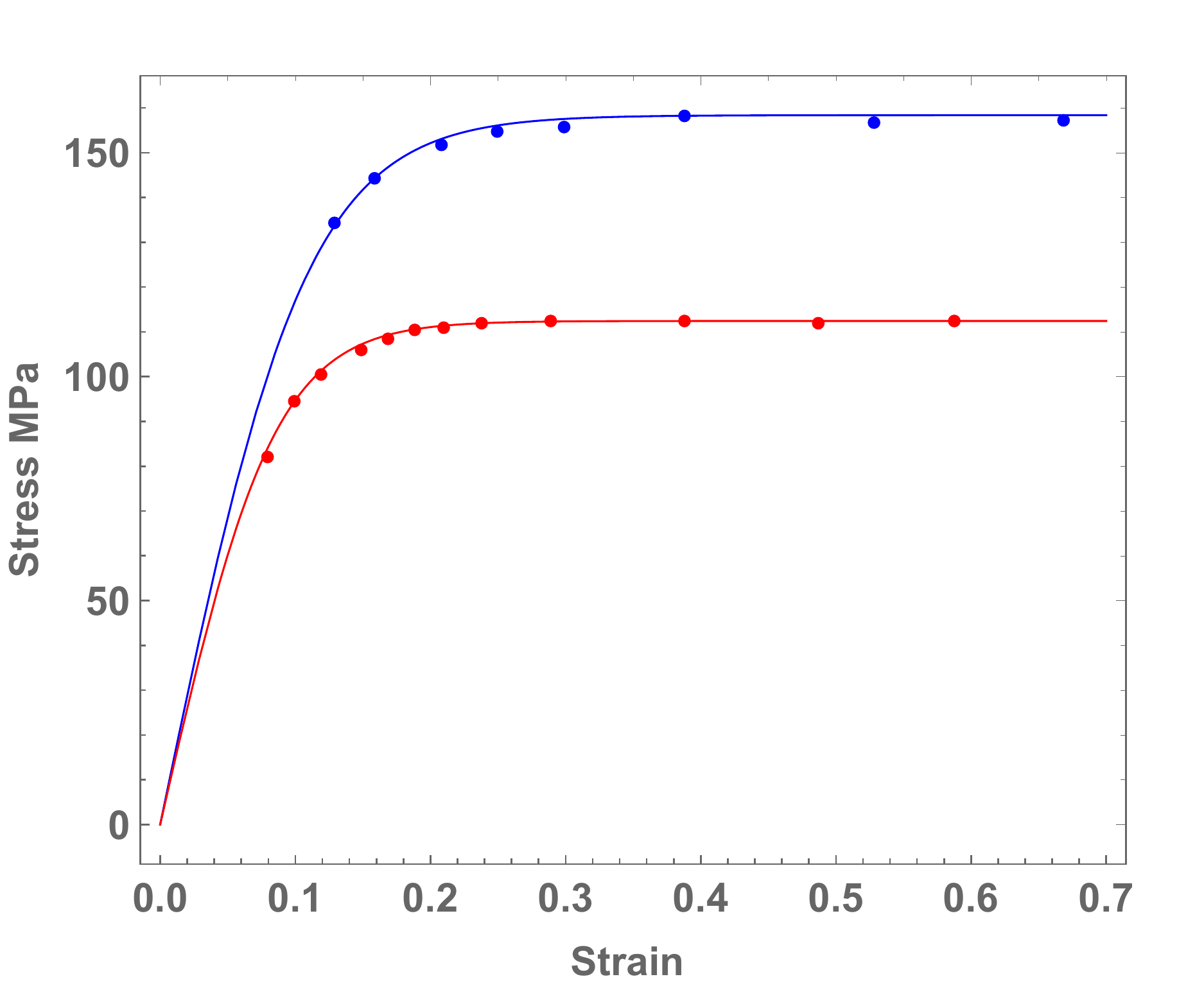}
\caption{Strain hardening curves for Cu: $T = 1023\,K$, $\dot\epsilon = 1,800/s$ (upper blue) and $T= 1173\,K$, $\dot\epsilon = 960/s$ (lower red).  The data points are from \cite{SAMANTA-71} }   \label{TDTFig-2}
 \end{center}
\end{figure}

Second is an equation of motion for $\rho$, which is a statement of energy conservation:
\begin{equation}
\label{dotrho} 
\dot\rho = \kappa_{\rho}\,{\sigma\,\dot\epsilon^{pl}\over \gamma_D}\,\Bigl[1- {\rho\over \rho_{ss}(\tilde\chi)}\Bigr].
\end{equation}
Here $\gamma_D$ is the dislocation energy per unit length, and $\kappa_{\rho}$ is the fraction of the input power $\sigma\,\dot\epsilon^{pl}$   that is converted into dislocations.  The second term inside the square brackets determines the rate at which dislocations are annihilated.  It does this by invoking a detailed-balance approximation using the effective temperature $\tilde\chi$; that is, it says that the density $\rho$ must approach the value given by Eq.(\ref{rhoss}), but with the steady-state $\tilde\chi_{ss}$ replaced by a time dependent $\tilde\chi$ during the approach to steady-state deformation.  

Finally, the equation of motion for $\tilde\chi$ is a statement of the first law of thermodynamics: 
\begin{equation}
\label{dotchi}
c_{e\!f\!f}\, e_D\,\dot{\tilde\chi}= \sigma\,\dot\epsilon^{pl}\,\Bigl( 1- {\tilde\chi\over\tilde\chi_{ss}}\Bigr) - \gamma_D\,\dot\rho,
\end{equation}
where $c_{e\!f\!f}$ is the effective specific heat. The second term in the parentheses is proportional to the rate at which effective heat is converted to ordinary heat, which reminds us that $\tilde\chi$ is a thermodynamically well-defined temperature. Like the comparable term in Eq.(\ref{dotrho}), this is a detailed-balance approximation.  The last term on the right-hand side accounts for energy stored in the form of dislocations.  

To illustrate the solutions of these equations of motion, we show in Fig. \ref{TDTFig-2}  just two of Samanta's 32 stress-strain data sets, compared here with the TDT predictions. The agreement between theory and experiment shown here is reassuringly excellent.  See \cite{KCL-20} for details about how the TDT equations were reformulated for numerical purposes and how parameter values were chosen for comparing their predictions with the experiments.  In computing the curves shown in Fig.\ref{TDTFig-2}, we simplified the analysis by neglecting Eq.(\ref{dotchi}) for $\tilde\chi$  and simply solving Eq.(\ref{dotrho}) with $\tilde\chi = \tilde\chi_{ss} = 0.23$, consistent with our observation in \cite{LBL-10} that $\tilde\chi \to \tilde\chi_{ss}$ very rapidly at high temperatures $T$. Our measured value of $\tilde\chi_{ss}$ is roughly consistent with our original guess that $\tilde\chi_{ss}\sim 0.25$.  The graphs in Fig. \ref{TDTFig-2} are almost identical to those shown  for wider ranges of temperatures and strain rates in the early TDT papers.  They also illustrate the invariance of the onset slopes for non-pre-hardened copper discovered experimentally by Kocks and Mecking \cite{KOCKS-MECKING-03} and explained  theoretically in \cite{LBL-10,JSL-17rev}.

We emphasize that these equations of motion are based entirely on  fundamental principles -- the laws of thermodynamics, energy conservation, and dimensional analysis.  Specific phenomena such as hardening, grain-size effects or yielding transitions play no role in deriving them.  Those phenomena are {\it predicted} by the equations.  The associated physical mechanisms are contained in the derivation of the double-exponential de-pinning formula, Eq.(\ref{doteps}),  and  in the conversion factors $\kappa_{\rho}$ and $c_{e\!f\!f}$ in Eqs. (\ref{dotrho}) and (\ref{dotchi}).  For example, the extreme stress sensitivity of the strain rate in Eq.(\ref{doteps}) naturally explains the $\rho$ and $T$ dependences of yield stresses; the phenomenological concept of a ``yield surface''  is unnecessary.  In a more specific way, the physically understandable grain-size dependence of the conversion factor $\kappa_{\rho}$ in Eq.(\ref{dotrho}) provides a simple explanation of Hall-Petch effects. Both of these predictions are discussed in \cite{JSL-17rev}.

One of the most remarkable aspects of these results is how many of the ingredients of conventional dislocation theory are completely absent in this elementary version of the theory.  The TDT dislocations are simply lines.  We do not ask whether they are edge dislocations or screw dislocations, or whether they are excess dislocations or geometrically necessary ones.  The crystals through which they move might be fcc, bcc, hcp, or something else.  Their motions are unaffected by crystalline orientations or slip planes or stacking faults.  They do not undergo cross slip.  They interact  with each other only at the pinning junctions and not via long-ranged elastic forces.  

Apparently we can go remarkably far with only this TDT caricature; but there must be limits. Finding and understanding those limits should be a high priority for new investigations.  Once we see what important physics is missing, we should be able to put realistic features back into the theory in fundamentally consistent ways, and thereby understand what roles they play and how important those roles may be. This process of making the TDT more realistic should help us distinguish useful phenomenological concepts from those that are unrealistic.  Our candidates for the latter category include distinctions between ``mobile'' and ``immobile'' dislocations, distinctions between different ``stages'' of strain hardening, and the idea that large flow stresses at high strain rates can be explained by something called ``phonon drag.'' At present, we see no scientific basis whatsoever for any of those conventional ideas.  

There is at least one limit to the validity of our scaling analysis.  We  have pointed out that the assumption of constant $\tilde\chi_{ss}$, and thus constant $\rho_{ss}$, must be changed at physically plausible, high strain rates.  Already, in \cite{LBL-10}, we showed how a simple strain-rate dependence of $\tilde\chi_{ss}$ with a corresponding increase in $\rho_{ss}$ can explain the high stresses observed in strong-shock experiments.  We thus found agreement between TDT and experiment over fifteen decades of strain rate.  This kind of analysis of high strain rates was also applied in \cite{JSL-18} to interpret molecular-dynamics simulations of crystalline deformation.  

There are other such issues, but most of them seem to be minor technicalities in comparison with a far more important question: What is the physics of brittle and ductile fracture in crystalline solids?  Basic theoretical research in this area has been at a decades-long standstill comparable to that which has afflicted theories of strain hardening.  

Consider the following. We know that solids are stronger when they are colder; their yield stresses and flow stresses increase with decreasing temperature.  This behavior is now  predicted by the TDT as seen in Eq.(\ref{doteps}) and its applications.  But we also know that solids become more brittle, i.e. they break more easily at lower temperatures despite the fact that they are stronger. How can these properties be consistent with each other? 

This basic question has not been answered.  So far as we know, it is not  even asked in the solid-mechanics literature.  The conventional model used for studying brittle or ductile crack initiation is one in which dislocations are emitted from infinitely sharp crack tips and move out along well defined slip planes. \cite{ARGON-01,TANAKAetal-08}  These dislocations either move freely, supposedly implying brittle behavior, or become dense enough to shield the crack tip and somehow produce ductility and toughness.  Agreement with experiment is modest at best.  As stated in a recent experimental paper by Ast et al. \cite{ASTetal-18}, an ``understanding of the controlling deformation mechanism is still lacking.'' Finding a  predictive theory of fracture toughness in crystalline solids should now be feasible, and should be a high priority for materials theorists.\\\\

\end{document}